\documentclass[conference]{IEEEtran}
\IEEEoverridecommandlockouts
\usepackage{eso-pic}
\usepackage{cite}
\usepackage{amsmath,amssymb,amsfonts}
\usepackage{graphicx}
\usepackage{textcomp}
\usepackage{xcolor}
\def\BibTeX{{\rm B\kern-.05em{\sc i\kern-.025em b}\kern-.08em
    T\kern-.1667em\lower.7ex\hbox{E}\kern-.125emX}}

\ifCLASSINFOpdf

\else

\fi
\usepackage{bm} 
\usepackage{siunitx}
\usepackage{times}  
\usepackage{helvet} 
\usepackage{courier}  
\usepackage[hyphens]{url}  
\usepackage{graphicx} 
\usepackage{graphicx}  
\usepackage[pagebackref=true,breaklinks=true,colorlinks,bookmarks=false]{hyperref}
\hypersetup{
    colorlinks=true,
    filecolor=magenta,      
    urlcolor=cyan,
}

\usepackage{subcaption}
\usepackage{amsmath,graphicx}

\usepackage{amsmath}
\usepackage{xcolor,soul,framed} 
\usepackage{xspace}
\usepackage{array}
\usepackage{eqparbox}
\usepackage{url}
\usepackage{longtable}
\usepackage{lipsum}
\usepackage{blindtext}
\usepackage{makecell}
\usepackage{mathtools}
\usepackage{commath}
\usepackage{multirow}
\usepackage{tabularx}
\usepackage[normalem]{ulem}
\usepackage{xspace}
\usepackage{algorithm}
\usepackage{algpseudocode}
\usepackage{amsfonts}
\usepackage{booktabs}
\usepackage{pifont}

\usepackage[table,xcdraw]{xcolor}

\newcolumntype{C}{>{\hsize=\dimexpr0.5\hsize+8\tabcolsep+\arrayrulewidth\centering\relax}X}

\newcommand{\cmark}{\ding{51}} 
\newcommand{\xmark}{\ding{55}} 

\begin{document}

\title{Ultrasound-Based Prediction of Cirrhosis Decompensation Using Large-Scale Computer Vision Models\\
{\footnotesize }
\thanks{$\dagger$These authors contributed equally to this work. *Corresponding author: asamir@mgh.harvard.edu.}
}

\author{
\IEEEauthorblockN{
Guangyi Zhang\textsuperscript{1,$\dagger$},
Peiyun Ni\textsuperscript{2,$\dagger$},
Eugene Cheah\textsuperscript{1},
Rajat Chandra\textsuperscript{3},
Peng Guo\textsuperscript{1}}

\IEEEauthorblockN{
Raymond T. Chung\textsuperscript{2},
Anthony E. Samir\textsuperscript{1, *}
}

\vspace{3mm}
\IEEEauthorblockA{\textsuperscript{1}
\textit{Center for Ultrasound Research $\&$ Translation, Massachusetts General Hospital, Harvard Medical School},
Boston, MA, USA
}
\IEEEauthorblockA{\textsuperscript{2}
\textit{Division of Gastroenterology, Massachusetts General Hospital, Harvard Medical School}, Boston, MA, USA
}
\IEEEauthorblockA{\textsuperscript{3}
\textit{Department of Medicine, Massachusetts General Hospital, Harvard Medical School}, Boston, MA, USA
}
}

\AddToShipoutPictureFG*{%
  \AtPageUpperLeft{%
    \raisebox{-0.30in}{%
      \makebox[\paperwidth][c]{%
        \footnotesize
        Accepted and presented at IEEE EMBC 2026.
        \textcopyright~2026 IEEE.
      }%
    }%
  }%
}

\AddToShipoutPictureFG*{%
  \AtPageLowerLeft{%
    \raisebox{0.35in}{%
      \makebox[\paperwidth][c]{%
        \parbox{0.8\paperwidth}{%
          \centering
          \fontsize{8}{9}\selectfont
          \textcopyright~2026 IEEE. Personal use of this material is permitted.
          Permission from IEEE must be obtained for all other uses,
          in any current or future media, including reprinting/republishing
          this material for advertising or promotional purposes, creating
          new collective works, for resale or redistribution to servers or
          lists, or reuse of any copyrighted component of this work in
          other works.
        }%
      }%
    }%
  }%
}
\maketitle

\begin{abstract}
Decompensation represents a critical transition in the course of cirrhosis, yet clinicians have limited non-invasive tools to reliably predict its onset. In this study, we propose a novel imaging-based approach that leverages large-scale computer vision models to analyze routine abdominal ultrasound images and extract predictive features beyond those captured by traditional laboratory-based risk scores. Ultrasound is widely available, low cost, and suitable for longitudinal surveillance, making it an attractive modality for scalable risk stratification and long-term follow-up. Our framework integrates automated ultrasound data processing with modern deep learning architectures to identify patients at high risk of decompensation prior to the occurrence of clinical deterioration. This non-invasive strategy offers a practical complement to existing clinical scoring systems and may enable earlier, more proactive management of patients with compensated cirrhosis.
\end{abstract}


\section{Introduction}
Cirrhosis is responsible for $1$ million deaths every year globally \cite{sepanlou2020global}, and is the 3\textsuperscript{rd} leading cause of death among adults $45$-$64$ years old \cite{asrani2019burden,gines2021liver}. In its natural course, cirrhosis progresses through two distinctive phases: compensated and decompensated cirrhosis. Patients in the compensated stage often appear clinically well, with a median survival of $12$ years \cite{d2006natural}. Over time, however, a rise in the blood pressure within the liver circulatory system due to cirrhosis (portal hypertension) can cause the progression into the decompensated stage. Decompensation is defined by the presence of any of the complications: abdominal fluid accumulation (ascites), gastrointestinal bleeding from enlarged veins (variceal bleeding), brain dysfunction due to liver failure (hepatic encephalopathy [HE]), kidney failure from liver dysfunction (hepatorenal syndrome [HRS]), or jaundice \cite{mansour2023british,d2022towards}. Patients with liver decompensation have a much lower median survival of around $2$ years \cite{d2006natural}. Given the drastic difference in survival between compensated and decompensated cirrhosis, early detection of this phase transition is critically important to enable targeted interventions.

Currently, the gold standard for assessing portal hypertension and the extent of liver dysfunction is hepatic venous pressure gradient (HVPG) measurement, which is invasive and involves catheterizing the jugular vein. Non-invasive prediction methods for liver decompensation remain limited, as existing laboratory-based scores are designed to predict risk of death \cite{child1964surgery,kamath2001model,kim2021meld}, but offer limited insights into decompensation risk. Meanwhile, liver stiffness measurements are typically validated only for specific types of cirrhosis, limiting their generalizability \cite{kim2012risk,thorhauge2024using}.

Ultrasound (US) is the most frequently used imaging modality in the longitudinal care of patients with cirrhosis, owing to its wide availability, low cost, portability, and absence of ionizing radiation. In contrast to cross-sectional imaging modalities such as computed tomography (CT) and magnetic resonance imaging (MRI), ultrasound is well suited for deployment in low-resource clinical settings \cite{rockey2009liver}. Routine ultrasound examinations capture rich information related to liver morphology, parenchymal texture, and portal hemodynamics, which may reflect early pathophysiological changes preceding clinical decompensation \cite{garcia2017portal,gines2021liver}. 
\textcolor{black}{Several prior studies have used abdominal ultrasound to predict liver-related outcomes in cirrhosis, using approaches ranging from semantic image assessment to radiomics and deep learning. In one study of patients with metabolic dysfunction-associated steatotic liver disease (MASLD), semantic ultrasound features did not outperform clinical laboratory variables in a random forest model predicting decompensation \cite{kosick2025machine}. In contrast, radiomic and deep learning features extracted from hepatic ultrasound images improved prediction of 30-day mortality in hepatitis B virus-related acute-on-chronic liver failure compared with models based on clinical variables alone \cite{huang2024ultrasound,huang2025predicting}. However, these prior studies focused on specific etiologic subgroups, leaving uncertainty about the generalizability of ultrasound-based models across broader compensated cirrhosis populations.}

\textcolor{black}{Transformer-based models have also shown promise in ultrasound-based medical imaging tasks, particularly for hepatocellular carcinoma (HCC) and MASLD. Vision transformer models have been used to extract ultrasound features for predicting microvascular invasion and early recurrence in patients with HCC \cite{chen2026ceus,zhang2025vit}, with some studies reporting improved performance compared with radiomic and clinical models \cite{qin2025transformer}. In MASLD, transformer-based ultrasound models have mainly been applied to diagnostic classification of steatosis and fibrosis \cite{bose2025deep}. However, none of these transformer-based models examined the risk of decompensation among patients with chronic liver disease.}
\textcolor{black}{To bridge this gap, our study introduces a non-invasive ultrasound-based framework for early prediction of liver decompensation by analyzing routinely acquired B-mode liver and spleen images using large-scale computer vision models and an automated real-world cohort processing pipeline. Our study evaluates both transformer-based and convolutional models for predicting future decompensation risks in patients with compensated cirrhosis across different etiologies.}

\textcolor{black}{Our contributions are threefold: 
(1) we propose a non-invasive framework based on routinely acquired ultrasound images from a clinical retrospective cohort for cirrhosis decompensation prediction; 
(2) we develop a robust data-processing pipeline that integrates optical character recognition (OCR)-based annotation extraction, Doppler image filtering, and clinician verification to ensure data quality and reproducibility; 
and (3) we conduct systematic benchmarking of $22$ model configurations across five architecture families under a unified evaluation protocol, enabling rigorous and transparent comparison of modern computer vision approaches.}

\section{Materials and Methods}
\subsection{Study Design}
This is a single-center, retrospective study approved by the Massachusetts General Brigham Institutional Review Board (IRB). The informed consent was waived.

\subsection{Study Population}
Our cohort includes $492$ adults with compensated cirrhosis and at least one abdominal US from the Mass General Brigham system ($1996$ to $2022$). We excluded studies that only contained Doppler or non-liver/spleen images after physician verification, reaching a final cohort of $486$ patients.

\begin{figure*}[h!]
    \begin{center}
    \includegraphics[width=2.0\columnwidth]{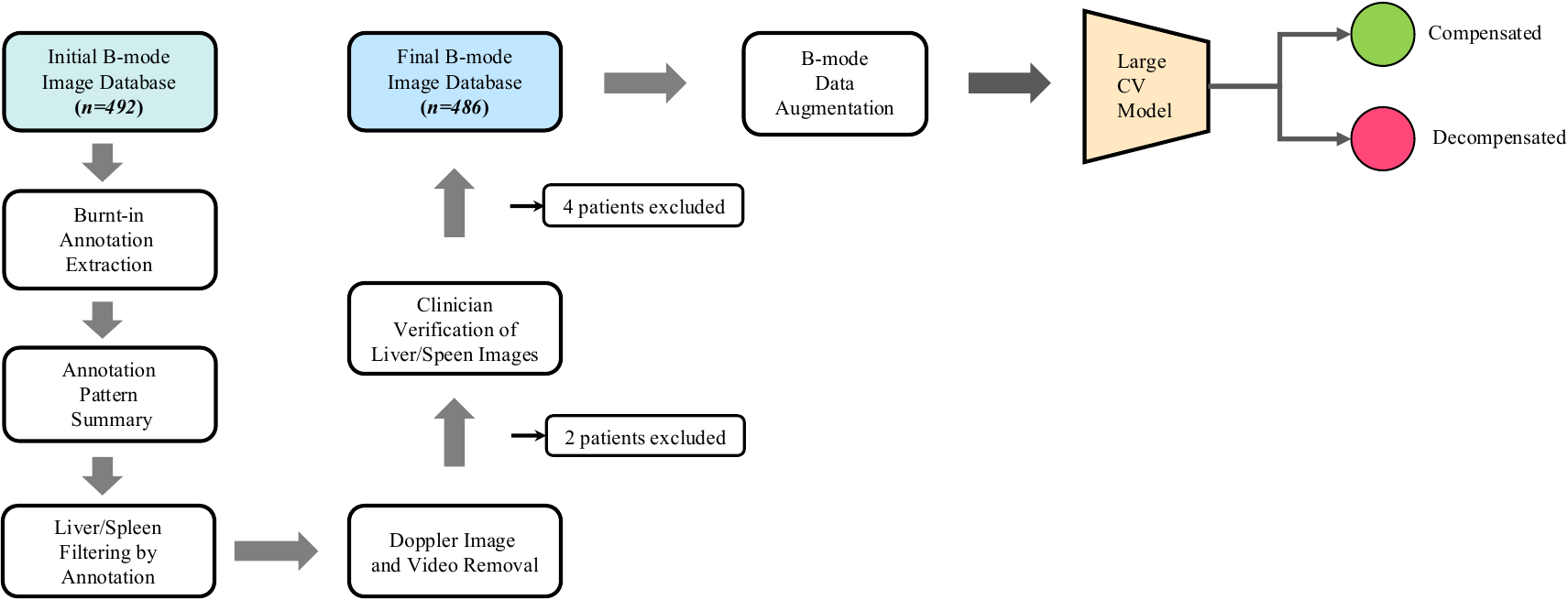} 
    \caption{B-mode data processing and modeling for cirrhosis decompensation prediction.}
    \label{pipeline}
    \end{center}
\end{figure*}

\subsection{Data Processing}
\textcolor{black}{A total of approximately $13{,}500$ ultrasound images, including both single-frame images and multi-frame acquisitions (raw Dicom files), were collected from $492$ patients using a diverse set of clinical ultrasound systems from multiple manufacturers, predominantly GE Healthcare (e.g., LOGIQ P5, LOGIQ E9, LOGIQ E10), GE Medical Systems (e.g., LOGIQ 9, LOGIQ 700), Siemens (e.g., Avanto, S1000, S2000) and few others, reflecting real-world clinical variability. For multi-frame acquisitions, only the first frame was used for subsequent analysis. This design choice was made to standardize input across single-frame and cine acquisitions, to avoid introducing strong temporal correlations among adjacent frames from the same examination. In addition, extracting all frames from cine acquisitions would disproportionately increase the number of highly correlated images contributed by examinations with long cine loops compared with those containing only a small number of static images, leading to patient-level imbalance and potential bias in model training and performance evaluation. For each patient, only one ultrasound examination from a single clinical visit was included since we aim to make predictions of cirrhosis decompensation for patients with compensated cirrhosis based on initial US around the time of cirrhosis diagnosis.} However, not all images were liver- or spleen-related, and a subset exhibited suboptimal image quality. To address these challenges, we developed a dedicated B-mode ultrasound data processing pipeline to systematically filter non-relevant images, perform quality control, and standardize inputs for downstream image learning.
The overview of data processing and modeling pipeline in shown in Figure \ref{pipeline}.
\subsubsection{Burnt-in Annotation Extraction}
Burnt-in text annotations were extracted from ultrasound DICOM frames using a dedicated optical character recognition (OCR) pipeline designed to identify organ-related labels embedded directly within the image. For each DICOM file containing pixel data, the pixel array was converted to a single 2D frame, accommodating grayscale, RGB, and cine formats by selecting the first frame when multi-frame data were present. Images were upsampled to $300\%$ of the original resolution using bicubic interpolation to improve OCR sensitivity to small on-screen text.

To improve robustness to vendor- and rendering-dependent appearance of burnt-in overlays, OCR was applied independently to multiple complementary image representations derived from each frame. Specifically, OCR was performed on the resized original image, an enhanced grayscale image, and each individual red, green, and blue (RGB) channel when color data were available. For grayscale and each RGB channel, images were intensity-inverted and enhanced using contrast-limited adaptive histogram equalization (CLAHE; clipLimit$=3.0$, tileGrid$=8\times8$) to increase contrast between dark text and bright backgrounds.

Text localization employed a region-scanning strategy that prioritizes areas where burnt-in overlays are most commonly displayed in ultrasound examinations. OCR was first applied to progressively larger bottom regions of the image, with crops spanning the bottom $20\%$-$50\%$ of the image height. If no valid text was detected in the bottom scans, OCR was subsequently applied to left-side regions, with crops covering $20\%$-$50\%$ of the image width. This bottom-first strategy with a left-side fallback rule reduced false detections from patient information and other non–organ-related metadata within the imaging field, while improving sensitivity to organ labels positioned near screen margins. OCR was performed using the Tesseract engine in LSTM-based recognition mode with a single-block text layout assumption, which is well suited for short, horizontally aligned burnt-in annotations.

Post-processing was applied to restrict detections to organ-related burnt-in annotations only. OCR outputs were filtered to retain uppercase alphabetic strings, excluding text containing lowercase characters, numerals, or special symbols. This filtering step removed non-informative overlays such as acquisition parameters, timestamps, scale markers, and machine settings, which are not relevant to organ identification. Frames for which no valid uppercase text was detected across all scanned regions and all image representations were considered negative and excluded from downstream analysis.

For each frame, OCR outputs from the original image, grayscale image, and individual RGB channels were retained independently and recorded in a structured format to support downstream quality control and organ-specific frame selection. This multi-channel, region-aware extraction strategy enabled robust identification of high-confidence organ-related burnt-in annotations while minimizing spurious detections from non-anatomical text.

\subsubsection{Annotation Pattern Summary}
To characterize the diversity and prevalence of organ-related burnt-in annotations identified by the OCR pipeline, we performed a keyword-guided aggregation of extracted text patterns across all processed frames. Following burnt-in annotation extraction, OCR outputs from the original image, grayscale image, and individual RGB channels were scanned for predefined liver- and spleen-related keywords, including anatomical organ names, lobe laterality terms, and commonly used clinical abbreviations (e.g., LIVER, LT/RT LOBE, RUQ/LUQ, PV, SPLEEN). Keyword matching was performed in a case-insensitive manner on a line-by-line basis.

For each frame, annotation detection was performed sequentially across image representations, and the process terminated as soon as a keyword match was identified. When a keyword-containing line was detected, uppercase alphabetic token sequences from that line were extracted and retained as a canonicalized \emph{annotation pattern}. Frames without any keyword matches across all representations were excluded from this summary. This early-exit strategy ensured efficient processing while prioritizing high-confidence organ-related annotations.

Extracted annotation patterns were aggregated across all frames to compute their frequencies, yielding a distribution of commonly observed views, organs, and anatomical descriptors in the dataset. This summary was necessary, as it enabled quantification of the coverage and variability of organ-related ultrasound views (e.g., sagittal and transverse orientations, left/right liver, and spleen) and supported the construction of a comprehensive, data-driven inventory of organ annotations, providing a compact representation of annotation heterogeneity across vendors and acquisitions and ensuring complete inclusion of organ-related images in downstream filtering.

\subsubsection{Liver/Spleen Filtering by Annotation}
Subsequently, filtering and refinement based on liver and/or spleen annotations were performed. OCR outputs extracted from multiple image representations (original, gray, red, green, blue) were scanned for the predefined set of liver and spleen related keywords, including organ names, laterality descriptors, and commonly used clinical abbreviations as mentioned above. Keyword matching was performed in a case-insensitive manner, and frames containing at least one keyword match in any OCR channel were retained, while frames without any matches across all channels were excluded.

To improve label reliability and suppress spurious detections, OCR outputs were further refined using a cross-channel consensus strategy. Candidate text lines were first filtered to retain uppercase alphabetic strings exceeding a minimum length and containing at least two words. High-confidence organ labels were identified through exact agreement across multiple OCR channels; when exact agreement was absent, a fuzzy matching criterion based on token-level substring overlap was applied, with preference given to longer and more consistent candidates. Remaining candidates lacking cross-channel support were subsequently verified by a clinician.

\subsubsection{Doppler Image Removal}
Doppler images were identified using a rule-based color pattern heuristic targeting the characteristic red–blue flow overlay. Images were first decomposed into RGB channels, and predominantly red pixels (high red intensity with relatively low blue and moderate green) and predominantly blue pixels (high blue intensity with relatively low red and moderate green) were detected. A frame was classified as Doppler if both red- and blue-dominant pixels exceeded a small area threshold (ratio $>0.005$ for each), indicating the presence of a red–blue color map consistent with Doppler flow visualization. Frames meeting this criterion were excluded, whereas the remaining frames were retained as non-Doppler (B-mode) candidates.

\subsubsection{Clinician Verification of Liver/Spleen Images}
Liver and spleen images were reviewed by a clinician to confirm anatomical relevance and remove residual false positives. This manual verification step served as a final quality control measure, ensuring the reliability of the automated filtering pipeline for downstream analysis.

\subsection{Final Cohort after Data Processing}
\begin{figure}[htbp]
\centering

\begin{subfigure}{\columnwidth}
    \centering
    \includegraphics[width=0.85\columnwidth]{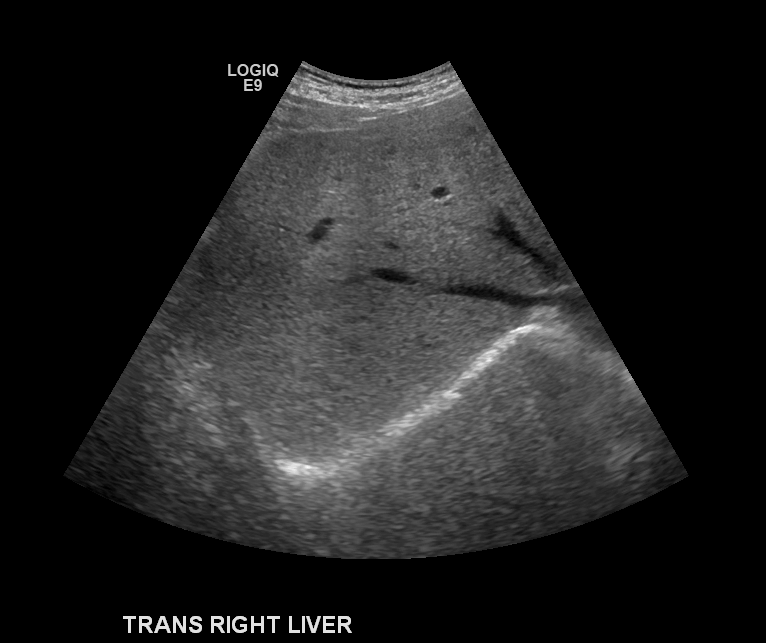}
    \caption{Liver B-mode ultrasound.}
\end{subfigure}

\vspace{0.5em}

\begin{subfigure}{\columnwidth}
    \centering
    \includegraphics[width=0.85\columnwidth]{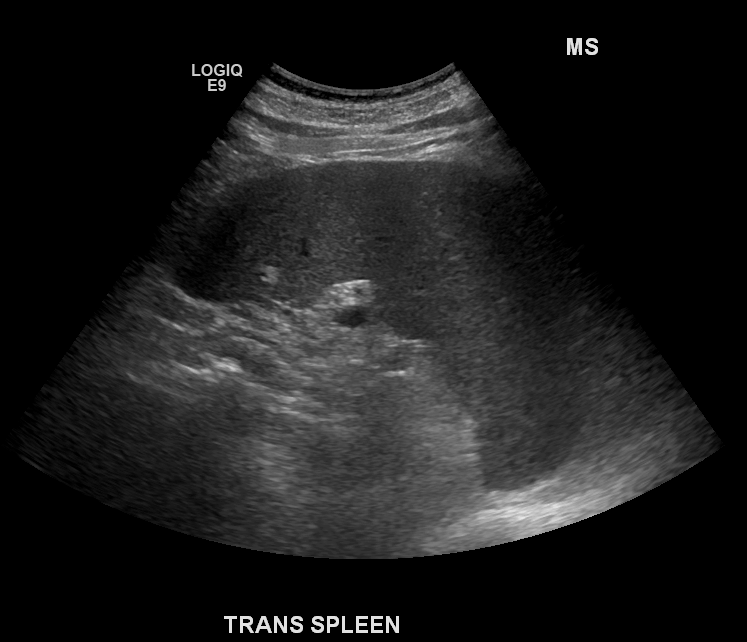}
    \caption{Spleen B-mode ultrasound.}
\end{subfigure}

\caption{\textcolor{black}{Representative B-mode ultrasound images of the liver and spleen.}}
\label{liver_spleen}
\end{figure}

Following data processing, $2$ patients were excluded because all available B-mode images were Doppler images, and $4$ patients were excluded after clinician verification due to false-positive annotation results, where no retained B-mode images contained liver or spleen anatomy. The final cohort comprised $7,840$ B-mode images from $486$ patients, where each image contains liver or/and spleen. On average, each patient contributed $16$ images, with $5,451$ images in the training set. The median age was $57$ years, and $60.4\%$ of patients were male. The predominant racial groups were White ($75.3\%$), Asian ($8.6\%$), and Black ($5.1\%$). Overall, $190$ patients ($39\%$) experienced decompensation over a median follow-up of $6.72$ years. \textcolor{black}{Examples of liver and spleen B-mode ultrasound images are shown in Figure \ref{liver_spleen}.}

\subsection{Modeling}
\subsubsection{B-mode Ultrasound Data Augmentation}
During training, B-mode ultrasound images were augmented by $10$-fold to improve robustness to acquisition variability while preserving anatomical fidelity. Mild geometric augmentation was applied, including horizontal flipping, small rotations ($\pm10^\circ$), slight scaling ($0.95$-$1.05$), and minor translations (up to $2\%$). Intensity augmentation included limited brightness and contrast perturbations. Low-level Gaussian noise and mild Gaussian blurring were applied to introduce controlled variability in image intensity and local spatial smoothness during training.

\subsubsection{Model Training and Architecture Evaluation}
Following data augmentation, we evaluated a comprehensive set of large computer vision models spanning multiple architectural families under a unified training protocol. The evaluated architectures included convolutional neural networks (ResNet \cite{he2016deep} and EfficientNet \cite{tan2019efficientnet}), transformer-based vision models (Vision Transformer (ViT) \cite{dosovitskiy2020image} and BEiT \cite{bao2021beit}), and hybrid transformer–convolutional models (MaxViT \cite{tu2022maxvit}). For each architecture family, base-scale and selected medium-to-large variants were evaluated to assess performance across different representational capacities. Models were trained using their standard input resolutions as defined by the architecture (e.g., $224$, $260$, $300$, and $384$), allowing assessment of resolution-dependent effects under a consistent training and evaluation protocol.

\subsection{Training and Evaluation Protocol}
We performed stratified splitting at the subject level using a fixed random seed. Specifically, $20\%$ of subjects were held out as an independent test set, and the remaining $80\%$ were used for model development and further split into training ($70\%$ of total subjects) and validation ($10\%$ of total subjects) sets. 
We ensured that no ultrasound images from the same patient overlapped across the training, validation, and test sets. 
Model selection was based on validation area under the receiver operating characteristic curve (AUROC) computed at each epoch, and the model checkpoint achieving the highest validation AUROC was saved and used for final evaluation on the independent test set. In addition, we report subject-wise (per-patient) AUROC, where subject-level predictions were obtained by averaging predicted probabilities across all images belonging to the same patient. In this study, because the outcome was defined as a binary decompensation event and no right-censoring was present at the evaluation time point, the concordance index (C-index) is mathematically equivalent to the AUROC. Accordingly, AUROC and C-index are used interchangeably throughout the manuscript.

All models were trained using PyTorch DistributedDataParallel (DDP) on a single node with $8$ NVIDIA V$100$ GPUs. Automatic mixed precision (AMP) with gradient scaling was employed to accelerate training and reduce memory usage. The per-GPU batch size was $64$, resulting in an effective batch size of $512$ across $8$ GPUs. Optimization was performed using AdamW (betas $0.9$ and $0.999$) with weight decay $0.05$. The base learning rate was set to $10^{-4}$ and linearly scaled by the number of GPUs, yielding an effective learning rate of $8\times10^{-4}$. A cosine annealing learning rate schedule was applied for $100$ epochs, with a minimum learning rate of $10^{-6}$. Training used cross-entropy loss with inverse-frequency class weighting and label smoothing ($0.1$) to mitigate class imbalance, and gradient clipping was applied with a maximum norm of $1.0$.

\section{Results}
We benchmark test-set performance against the Model for End-Stage Liver Disease-Sodium (MELD-Na) score to maintain complete data availability across subjects.
MELD-Na is a clinic score calculated based on laboratory test results of serum creatinine, total bilirubin, international normalized ratio (INR), and serum sodium. It is a validated clinical score to predict mortality in patients with cirrhosis and to prioritize organ allocation for liver transplant \cite{biggins2006evidence}.

\begin{table}[h]
\centering
\caption{Concordance index (C-index) with 95\% confidence intervals for imaging-based survival models. Models are grouped by architecture family and ordered with training from scratch shown before ImageNet pretraining. The benchmark serum-based method, MELD-Na, is included for comparison.}
\label{tab:subject_C-index}
\renewcommand{\arraystretch}{1.15}
\begin{tabular}{llcc}
\toprule
\textbf{Family} & \textbf{Architecture} & \textbf{Pretrained} & \textbf{C-index (95\% CI)} \\
\midrule
\multirow{4}{*}{ResNet}
 & ResNet-50  & \xmark & 0.58 (0.46--0.69) \\
 & ResNet-50  & \cmark & \underline{0.69} (0.58--0.79) \\
 & ResNet-101 & \xmark & 0.59 (0.46--0.70) \\
 & ResNet-101 & \cmark & 0.64 (0.52--0.75) \\
\midrule
\multirow{6}{*}{EfficientNet}
 & EfficientNet-B2 & \xmark & 0.59 (0.47--0.71) \\
 & EfficientNet-B2 & \cmark & 0.59 (0.47--0.70) \\
 & EfficientNet-B3 & \xmark & 0.55 (0.44--0.67) \\
 & EfficientNet-B3 & \cmark & 0.65 (0.53--0.76) \\
 & EfficientNet-B4 & \xmark & 0.56 (0.44--0.68) \\
 & EfficientNet-B4 & \cmark & 0.60 (0.48--0.71) \\
\midrule
\multirow{4}{*}{BEiT}
 & BEiT-Base-224 & \xmark & 0.52 (0.40--0.64) \\
 & BEiT-Base-224 & \cmark & 0.56 (0.45--0.68) \\
 & BEiT-Base-384 & \xmark & 0.56 (0.45--0.68) \\
 & BEiT-Base-384 & \cmark & 0.54 (0.42--0.67) \\
\midrule
\multirow{4}{*}{ViT}
 & ViT-Base-224 & \xmark & 0.53 (0.41--0.65) \\
 & ViT-Base-224 & \cmark & 0.58 (0.46--0.70) \\
 & ViT-Base-384 & \xmark & 0.53 (0.41--0.65) \\
 & ViT-Base-384 & \cmark & \textbf{0.70} (0.58--0.81) \\
\midrule
\multirow{4}{*}{MaxViT}
 & MaxViT-Base  & \xmark & 0.61 (0.50--0.72) \\
 & MaxViT-Base  & \cmark & 0.68 (0.57--0.78) \\
 & MaxViT-Small & \xmark & 0.57 (0.46--0.69) \\
 & MaxViT-Small & \cmark & 0.62 (0.50--0.73) \\
 \midrule
  \midrule
 MELD-Na & -& - & 0.61 (0.50--0.72) \\
\bottomrule
\end{tabular}
\vspace{1mm}
\captionsetup{font=small}
\caption*{Note: The highest C-index is shown in bold, and the second-highest C-index is shown with an underline.}
\end{table}

As shown in Table~\ref{tab:subject_C-index}, we compared the best-performing B-mode ultrasound learning models with the MELD-Na score. The model using liver and spleen ultrasound features achieved strong predictive performance for cirrhosis decompensation, with the highest C-index of $0.70$ ($95\%$ confidence interval (CI): $0.58$--$0.81$), obtained using a pretrained Vision Transformer (ViT-Base-384). This performance was closely followed by a pretrained ResNet-50 model, which achieved a C-index of $0.69$ ($95\%$ CI: $0.58$--$0.79$). Both ultrasound-based models had higher point estimates for the C-index than MELD-Na ($0.61$ [$95\%$ CI: $0.50$--$0.72$]), suggesting the potential for further performance gains with model refinement.
\textcolor{black}{Paired DeLong's test did not identify a statistically significant difference in AUROC between the pretrained ViT-Base-384 model and MELD-Na ($p=0.31$), potentially due to the relatively small test cohort size ($n=98$).}

We also evaluate the impact of pretraining, across the $n=11$ architecture configurations for which both pretrained and randomly initialized models were available, pretraining consistently improved subject-level discriminative performance. The mean/median increase in C-index was $\Delta=+0.06/+0.05$, indicating that performance gains were not driven by a small number of outliers but were broadly observed across models.
A paired Wilcoxon signed-rank test demonstrated that the improvement associated with pretraining was statistically significant ($p=0.0029$). These results suggest that transferring representations learned from large-scale natural image datasets substantially enhances model generalization for ultrasound-based risk prediction, even in the presence of domain differences between natural images and medical ultrasound.

\section{Discussion}
\textcolor{black}{This study demonstrates the potential of using routinely acquired liver and spleen B-mode ultrasound images to predict future decompensation risk in patients with compensated cirrhosis. By learning directly from real-world ultrasound images, the proposed deep learning approach may capture prognostic imaging patterns beyond conventional laboratory-based scores.}

\textcolor{black}{The improvement observed with pretrained models suggests that transfer learning may be useful for ultrasound-based prognostic modeling, especially in moderate-sized medical imaging cohorts. However, given the wide confidence intervals, these findings should be interpreted as preliminary support for this imaging-based approach rather than definitive superiority over MELD-Na.}

\textcolor{black}{This study also has several limitations. First, only the baseline ultrasound examination was analyzed because follow-up imaging was inconsistent across patients. Second, elastography was excluded due to limited availability, with data present in fewer than $20\%$ of patients. Third, the single-system cohort of 486 patients may limit generalizability.}

\section{Conclusion}
This study demonstrates that a rigorously designed ultrasound data processing pipeline, combined with large-scale computer vision models, has the potential to accurately predict cirrhosis decompensation from B-mode ultrasound images. Leveraging automated burnt-in annotation extraction, organ-specific filtering, and robust quality control, the proposed framework identifies clinically meaningful liver and spleen features without reliance on manual region of interest placement. Across the evaluated architectures, models trained on processed ultrasound images achieved promising subject-level discrimination. The best-performing pretrained vision transformer obtained a higher point estimate of the C-index than the MELD-Na score based on laboratory results. These findings highlight the potential of modern computer vision approaches to complement traditional clinical risk models for decompensation risk stratification. Future validation in larger, multi-center cohorts will be important to assess generalizability and clinical utility.
\textcolor{black}{Future work will also incorporate explainability methods to identify image regions contributing to predictions and enhance clinical interpretability.}

\bibliographystyle{IEEEtran}
\bibliography{IEEEabrv,refs}

\vspace{12pt}

\end{document}